\documentclass[twocolumn,showpacs,preprintnumbers,amssymb,aps,pra]{revtex4}

\usepackage{graphicx}
\usepackage{dcolumn}
\usepackage{bm}
\usepackage{latexsym,epsfig}

\usepackage{array}
\usepackage{color}
\usepackage{ulem}
\usepackage{amstext}

\begin{document} 

\title{Sympathetic cooling of $^{113}$Cd$^+$ by laser-cooled $^{40}$Ca$^+$ in a linear Paul trap for Microwave Ion Clocks}
\vspace*{0.5cm}

\author{J. Z. Han$^{1,2}$, H. R. Qin$^{1,2}$, L. M. Guo$^{1}$, N. C. Xin$^{1}$, H. X. Hu$^{1}$, Y. M. Yu$^{3}$, V. A. Dzuba$^4$\footnote{Email: v.dzuba@unsw.edu.au}, J. W. Zhang$^{1}$\footnote{Email: zhangjw@tsinghua.edu.cn}, and L. J. Wang$^{1,2}$\footnote{Email: lwan@mail.tsinghua.edu.cn}}

\affiliation{$^1$State Key Laboratory of Precision Measurement Technology and Instruments, Department of Precision Instruments, Tsinghua University, Beijing 100084, China}
\affiliation{$^2$Department of Physics, Tsinghua University, Beijing 100084, China}
\affiliation{$^3$Beijing National Laboratory for Condensed Matter Physics, Institute of Physics, Chinese Academy of Sciences, Beijing 100190, China}
\affiliation{$^4$School of Physics, University of New South Wales, Sydney, 2052, Australia}

\date{Received date; Accepted date}

\pacs{}

\vskip1.0cm

\begin{abstract}
We report sympathetic cooling of $^{113}$Cd$^+$ by laser-cooled $^{40}$Ca$^+$ in a linear Paul trap for microwave clocks. Long-term low-temperature confinement of $^{113}$Cd$^+$ ions was achieved. The temperature of these ions was measured at $90(10)$ mK, and the corresponding uncertainty arising from the second-order Doppler shifts was estimated to a level of $2\times10^{-17}$. Up to $4.2\times10^5$ Cd$^+$ ions were confined in the trap, and the confinement time constant was measured to be 84 hours. After three hours of confinement, there were still $10^5$ Cd$^+$ ions present, indicating that this Ca$^+$--Cd$^+$ dual ion system is surprisingly stable. The ac Stark shift was induced by the Ca$^+$ lasers and fluorescence, which was carefully estimated to an accuracy of $5.4(0.5)\times10^{-17}$ using a high-accuracy \textit{ab initio} approach. The Dick-effect-limited Allan deviation was also deduced because deadtimes were shorter. These results indicate that a microwave clock based on this sympathetic cooling scheme holds promise in providing ultra-high frequency accuracy and stability.
\end{abstract}

\maketitle

\section{Introduction}
With the development of atomic clock technology, atomic clocks have played an important role in both basic physics \cite{Dzuba-NP-2016,Safronova-RMP-2018,Wcislo-NA-2016} and practical applications \cite{Hinkley-science-2013}. Because of their long-term stability and relatively simple structure, microwave clocks play an irreplaceable role in areas such as satellite navigation \cite{Bandi-EL-2011}, deep space exploration \cite{Prestage-IEEE-2007,Tjoelker-PTTI-2012}, and time keeping \cite{Diddams-science-2004}. To date, cesium fountain clocks have reigned supreme in state-of-the-art performance regarding frequency uncertainty. Uncertainties have been reduced to the $10^{-16}$ level, the main limitation being collision shifts, black-body radiation (BBR) shifts, and cavity phase shifts \cite{Weyers-Metrologia-2018}. Among microwave atomic clocks, those based on ions have attracted extensive research because of unique advantages. First, ion clocks have small collision shifts because of the strong Coulomb repulsion between ions. Second, ions confined in the trap occupy a small space volume, making it easier to measure the temperature with high precision by reducing the uncertainty associated with BBR shifts. Third, ion clocks need no microwave cavity, and therefore uncertainties from cavity phase shifts do not arise. In fountain clocks, such shifts arise from the asymmetry of the electric fields in the microwave cavity when atoms move up and down.

To date, $^{113}$Cd$^+$ \cite{Jelenkovic-PRA-2006,KMiao-OL-2015}, $^{171}$Yb$^+$ \cite{Phoonthong-APB-2014,Mulholland-APB-2019}, and $^{199}$Hg$^+$ \cite{Berkeland-PRL-1998,Burt-ITUFFC-2016} ions have been used in the development of microwave ion clocks. A great deal of progress has been achieved. In the past ten years, our group has been committed to developing a microwave atomic clock based on $^{113}$Cd$^+$ \cite{KMiao-OL-2015,JWZhang-PRA-2012,SGWang-OE-2013,JZHan-EPJD-2019,YNZuo-APL-2019,JZHan-PRA-2019}. At present, the short-term frequency instability has reached $6.1\times10^{-13}/\sqrt{\tau}$, and the uncertainty associated with ground-state hyperfine splitting has reached $6.6\times10^{-14}$ \cite{KMiao-OL-2015}. Currently, the fundamental limitation halting further improvements in accuracy and stability of microwave ion clocks are the second-order Doppler shifts (SODSs), and the Dick effect resulting from dead times during interrogations \cite{JWZhang-CPL-2015}.

In the past few years, sympathetic-cooling technology has greatly advanced the development of the optical ion clock. Optical ion clocks based on Mg$^+$ sympathetic cooling of Al$^+$ have become one of the most accurate atomic clocks in the world \cite{Brewer-PRL-2019}. This cooling technology cools the target ions continuously to a lower temperature through Coulomb interactions with the coolant ion. Sympathetic cooling could be an effective method to overcome the limitations on stability and accuracy of microwave ion clocks. 
The application of sympathetic cooling to microwave ion clocks was earlier proposed by D. J. Wineland and collaborators from National Institute of Standards and Technology, and a preliminary scheme was studied involving the use of $^{198}$Hg$^+$ \cite{Larson-PRL-1986} and $^{26}$Mg$^+$ \cite{Bollinger-IEEE-1991} sympathetic cooling $^9$Be$^+$ in further developments of the Be$^+$ frequency standard in the Penning trap. 
Microwave ion clocks need to trap a large number of ions stably to improve the signal-to-noise ratio (SNR) during the interrogation of clock signals. Under the sympathetic cooling scheme, Coulomb repulsion of the inner shells of ions makes the stable trapping of large numbers of dual ion systems very challenging experimentally. With the cooling lasers of the coolant ions operating during microwave interrogations, the light shifts caused by these lasers and the fluorescence of the coolant ions must be evaluated carefully.

We had earlier proposed using $^{24}$Mg$^+$ to sympathetically cool $^{113}$Cd$^+$ and have reported results of our experiment \cite{YNZuo-APL-2019}. Compared with that scheme, using $^{40}$Ca$^+$ as coolant is more promising because of certain key advantages. First, the mass ratio of $^{113}$Cd$^+$ and $^{40}$Ca$^+$ is smaller than $^{113}$Cd$^+$ and $^{24}$Mg$^+$ and therefore increases the cooling efficiency and improves the trapping stability. Second, the reaction rate of $^{40}$Ca$^+$ with residual background H$_2$ are much smaller than $^{24}$Mg$^+$. The reaction turns the Mg$^+$ into dark ions and is sufficient to limit the cooling efficiency and stability of confinement. Third, the lasers for $^{40}$Ca$^+$ are much easier to obtain than those for $^{24}$Mg$^+$. All the lasers needed for $^{40}$Ca$^+$ cooling can be generated by diode lasers, apart from the ultraviolet laser for $^{24}$Mg$^+$ which must be obtained by a fourth-harmonic-generation laser system. The laser system implemented for $^{40}$Ca$^+$ has better reliability and is cost effective compared with that of $^{24}$Mg$^+$. 

In this study, a large number of $^{113}$Cd$^+$ ions sympathetically cooled by $^{40}$Ca$^+$ and stably confined in a linear Paul trap was experimentally achieved. The number of $^{113}$Cd$^+$ ions and $^{40}$Ca$^+$ ions reached were $4.2\times10^5$ and $1.7\times10^5$, respectively, and the confinement time constant reached was 84 hours. The temperature of the Cd$^+$ ions was as low as $90(10)$ mK , and the corresponding SODS was estimated to be $-1.84(0.2)\times10^{-16}$. The ac Stark shifts caused by the coolant ions were carefully estimated by using an accurate \textit{ab initio} approach; the magnitude and uncertainty at wavelengths 397 nm and 866 nm were $5.4(0.5)\times10^{-17}$ and $2.0(0.2)\times10^{-19}$, respectively. The Dick-effect-limited Allan deviation was also deduced as there was no need for an extra cooling procedure; its value was decreased to $1.8\times10^{-14}/\sqrt{\tau}$. These results show that using Ca$^+$ as coolant ions to sympathetically cool Cd$^+$ is very stable and efficient. The main uncertainty of the systematic frequency shifts was significantly reduced to $10^{-16}$ level.

\section{Experiment Scheme}
The entire experimental system consists of three parts: the ion trap, the lasers, and the detection system. The ions are trapped in a linear Paul trap using four electrodes. Every electrode is separated into three segments. The middle acts as radio frequency (RF) electrodes for radial confinement, and the ends act as endcap (EC) electrodes for axial confinement. The driving frequency chosen is 2.076 MHz, and the RF voltage varies from 0 V to 500 V, which covers the common stable areas for $^{113}$Cd$^+$ and $^{40}$Ca$^+$ ions. The EC voltage varies from 0 V to 100 V. To reduce RF heating effects and increase the number of trapped ions, the ratio of $R/r_0$ was optimized and set at 1.1468 \cite{Denison-JVST-1971}, where $2R=14.22$ mm is the outer diameter of the electrode and $r_0=6.2$ mm the radial distance from the axis of the trap to the closest surface of the electrodes. The vacuum of the vacuum chamber is maintained by an ion pump, the background pressure being around $5\times10^{-10}$ mBar. Three pairs of Helmholtz coils are used to generate a static magnetic field to split the Zeeman sublevels and compensate for the geomagnetic field.

\begin{figure}
\centering
\resizebox{0.4\textwidth}{!}{
\includegraphics{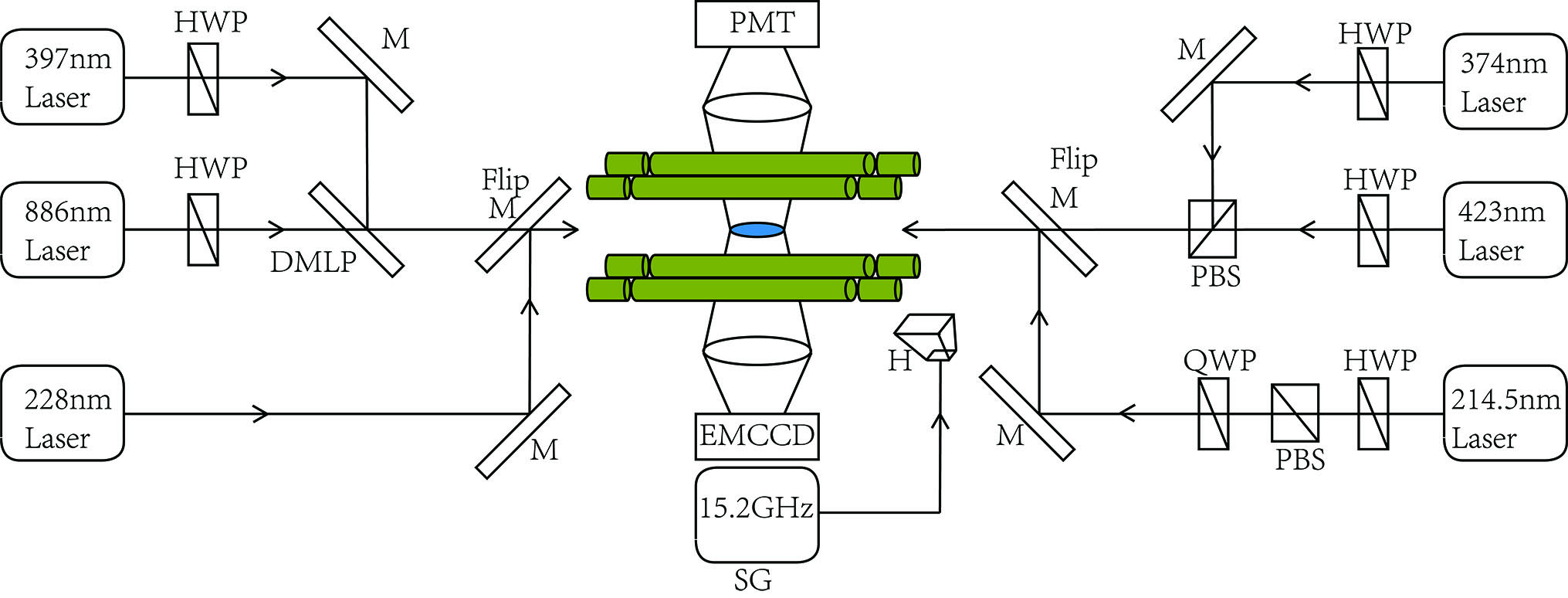}
}
\caption{Schematic diagram of the experimental setup. M: mirror; Flip M: mirror with flipper; DMLP: long pass dichroic mirror; HWP: half-wave plate; QWP: quarter-wave plate; PBS: polarizing beam splitter; PMT: photomultiplier tube; EMCCD: electron-multiplying charge-coupled device; SG: signal generator; H: horn antenna. A static magnetic field, B, is applied and its direction is parallel to the trap electrodes.}
\label{fig:guanglu} 
\end{figure}

\begin{figure}
\centering
\resizebox{0.4\textwidth}{!}{
\includegraphics{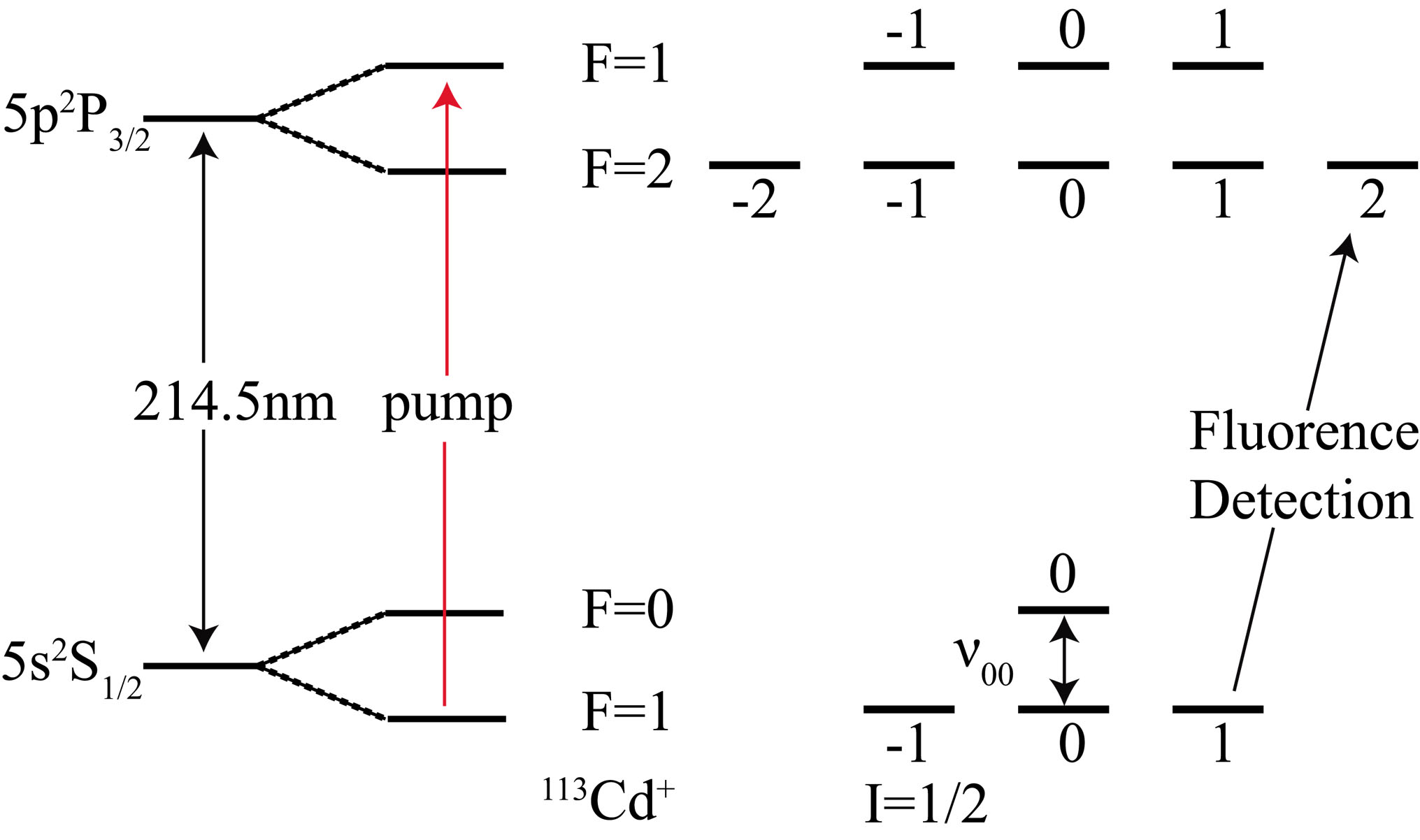}
}
\caption{Energy level scheme of $^{113}$Cd$^+$(Not to scale).}
\label{fig:level} 
\end{figure}

We chose photoionization instead of electron bombardment ionization to reduce disturbances from stray charges. For the laser setup of the experiment (Fig.~\ref{fig:guanglu}), the 423-nm (Ca I $4s^2\ ^1S_0\rightarrow4s4p\ ^1P_1$) laser and the 374-nm lasers are used to ionize neutral $^{40}$Ca into singly charged ions $^{40}$Ca$^+$. The 397-nm (Ca II $4s\ ^2S_{1/2}\rightarrow4p\ ^2P_{1/2}$) laser and the 866-nm (Ca II $3d\ ^2D_{3/2}\rightarrow4p\ ^2P_{1/2}$) laser are used for Doppler cooling and repumping the $^{40}$Ca$^+$ ions. For $^{113}$Cd$^+$, we use the 228-nm (Cd I $5s^2\ ^1S_0\rightarrow5s5p\ ^1P_1$) laser for two-photon ionization of neutral $^{113}$Cd. Because of the hyperfine structure of $^{113}$Cd$^+$, excited dark states are generated. To improve the cooling and fluorescence detection efficiency, we use a circular polarization laser beam to connect two Zeeman sublevels $5s~^2S_{1/2}$ (F=1 m$_F$=1) $\rightarrow$ $5p~^2P_{3/2}$ (F=2 m$_F$=2) to prevent ions from transitioning into dark states $5p~^2P_{3/2}$ (F=1), and a 15.2-GHz microwave RF signal to repump the ions in dark states $5s~^2S_{1/2}$ (F=0) back to the cycling transition state. The energy levels of $^{113}$Cd$^+$ are shown in Fig.~\ref{fig:level}.

For the detection system, we use an electron-multiplying charge-coupled device (EMCCD) to image the structure of the dual-ion system, and a photomultiplier tube (PMT) to determine the number of ion-scattered photons quantitatively. The lens between the EMCCD and PMT are well designed to reduce aberration. The wavelengths used for detection are 397 nm and 214.5 nm for $^{40}$Ca$^+$ and $^{113}$Cd$^+$, respectively.

\section{Temperature and Ion-loss Measurements}

To apply a sympathetic cooling scheme to a microwave ion clock, two conditions must be met simultaneously: i) a large number of target ions must be stably trapped in the ion trap, and ii) a high sympathetic cooling efficiency must be maintained to keep the target ions at a low temperature.

\begin{figure}
\centering
\resizebox{0.5\textwidth}{!}{
\includegraphics{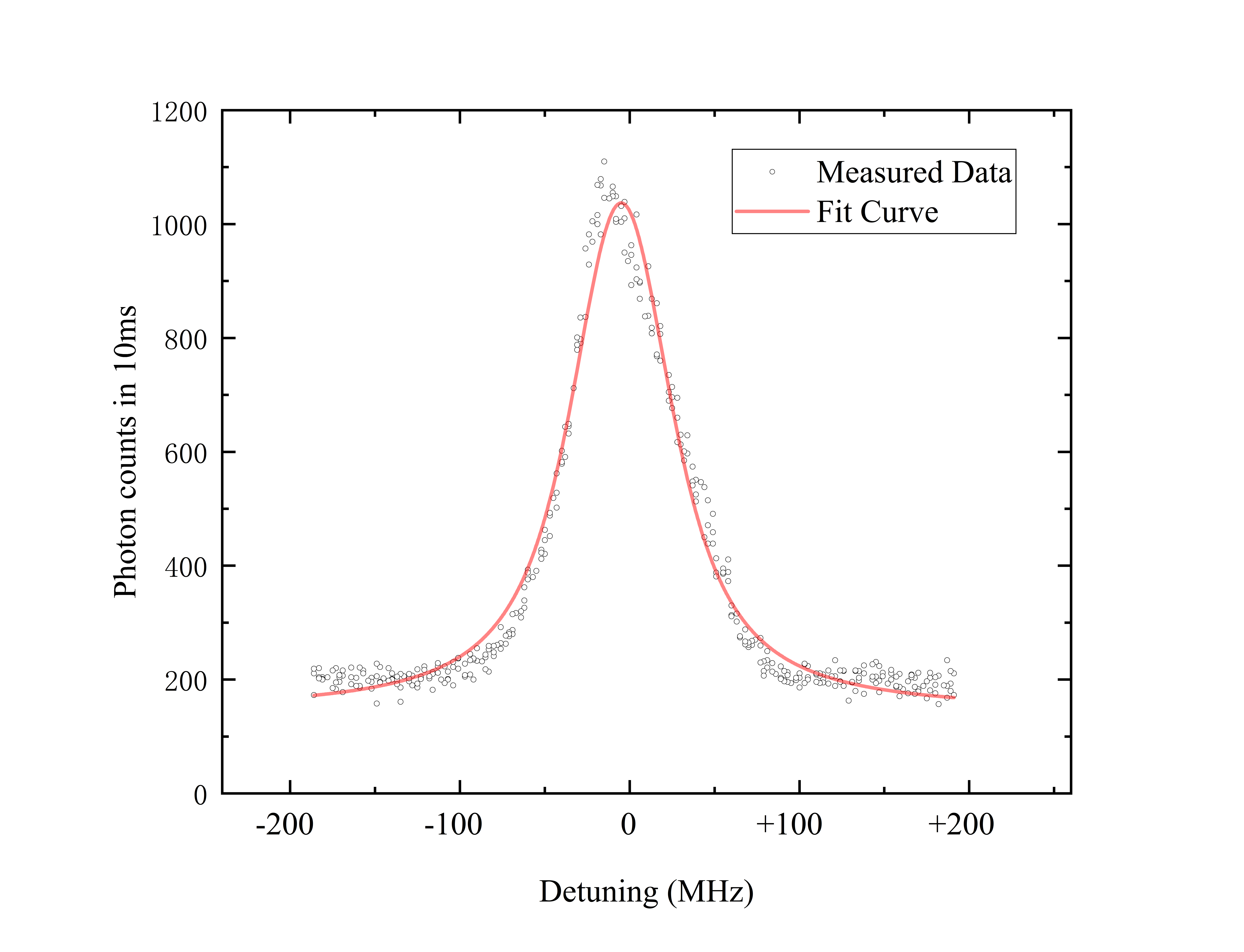}
}
\caption{Typical temperature measurement obtained using a Voigt fitting profile.}
\label{fig:89mk} 
\end{figure}

The measurement of the ion temperature is achieved by measuring Doppler broadening of the laser cooling transition \cite{Larson-PRL-1986,Bollinger-PRL-1984}. Fig.~\ref{fig:89mk} is a typical measurement result. To reduce cooling and heating effects of the detection laser, the laser power is maintained below 20$\rm{\mu}$W. The fitting curve is a Voigt profile, expressed as \cite{origin}
\begin{eqnarray}
F&=&F_0+(F_L\ast F_G)(\nu),\nonumber \\
F_L(\nu)&=&\frac{2A}{\pi}
\frac{\omega_L}{4(\nu-\nu_c)^2+\omega_L^2},\nonumber \\
F_G(\nu)&=&\sqrt{\frac{4\ln2}{\pi}}
\frac{e^{-\frac{4\ln2}{\omega^2_G}*\nu^2}}{\omega_G},
\end{eqnarray}
where $F_0$ denotes the offset, $\nu$ the laser frequency, $\nu_c$ the laser center frequency, $A$ the area, $\omega_L$ the Lorentzian width, $\omega_G$ the Gaussian width of Doppler broadening, and $*$ means convolution. For $^{113}$Cd$^+$, $\omega_L=60.13$ MHz, which is the natural linewidth of the D$_2$ transition of $^{113}$Cd$^+$; the fitted $\omega_G=28.68(1.4)$ MHz. The ion temperature is then determined using \cite{YNZuo-APL-2019,Fritz-book-2006}
\begin{equation}
T=\frac{Mc^2}{8\ln2~k_B}(\frac{\omega_G}{\nu_c})^2,
\end{equation} 
where $M$ denotes the mass of $^{113}$Cd$^+$, $c$ the speed of light, and $k_B$ the Boltzmann constant. The corresponding ion temperature is $93(1.4)$ mK, which is the upper limit to the temperature of the ions.

\begin{figure}
\centering
\resizebox{0.5\textwidth}{!}{
\includegraphics{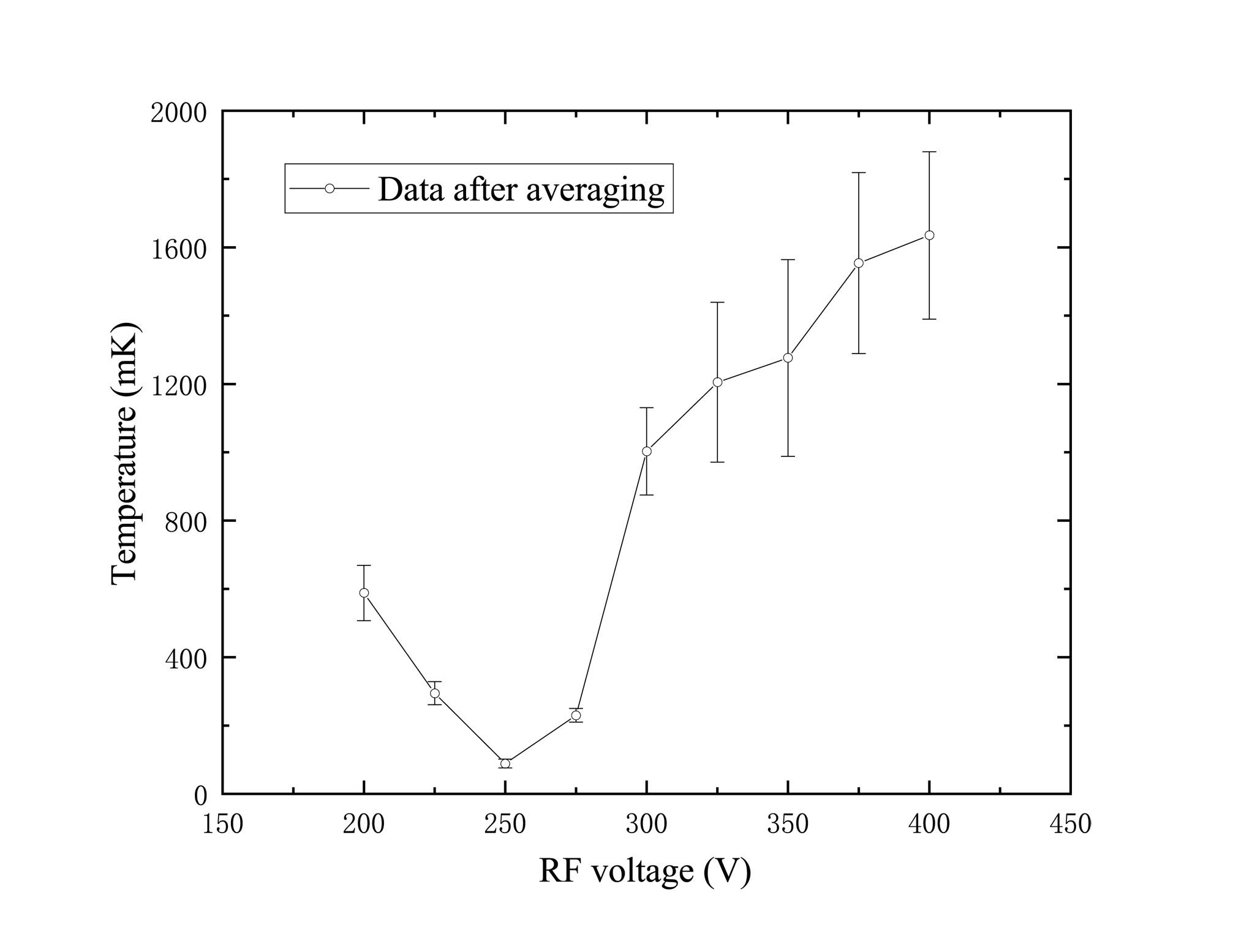}
}
\caption{RF voltage affects the sympathetic cooling efficiency; the endcap voltage is fixed at 10 V while changing the RF voltage. Each experimental data point is the average of three measurements.}
\label{fig:rf} 
\end{figure}

Although we have optimized the experimental setup for the dual-ion system, we still need to find out the best experimental parameter settings to achieve stable trapping of a large number of ions in the dual-ion system and efficient cooling of target ions. The main heating source stems from RF heating effects, and hence we measured with different RF voltages the temperature of the $^{113}$Cd$^+$ ions sympathetically cooled by $^{40}$Ca$^+$. From the results (Fig.~\ref{fig:rf}), we see there is a positive correlation between the ion temperature and the RF voltage, signifying RF heating strongly influences ion temperatures. When the RF voltage is above 300 V, we see the ion temperature rising significantly with increasing RF voltage. These hot $^{40}$Ca$^+$ ions cannot be maintained in the liquid phase, and the cloud states cannot cool the $^{113}$Cd$^+$ ions to very low temperatures. When the RF voltage is below 250 V, the temperature increases with decreasing RF voltage which we attribute to a decrease in the trap depth. The edge of the RF trap potential may contribute additional heating effects through imperfections in the potential wells. When the RF potential is below 200 V, the $^{113}$Cd$^+$ ions cannot be stably confined, and the number of ions decreases rapidly.

\begin{figure}
\centering
\resizebox{0.5\textwidth}{!}{
\includegraphics{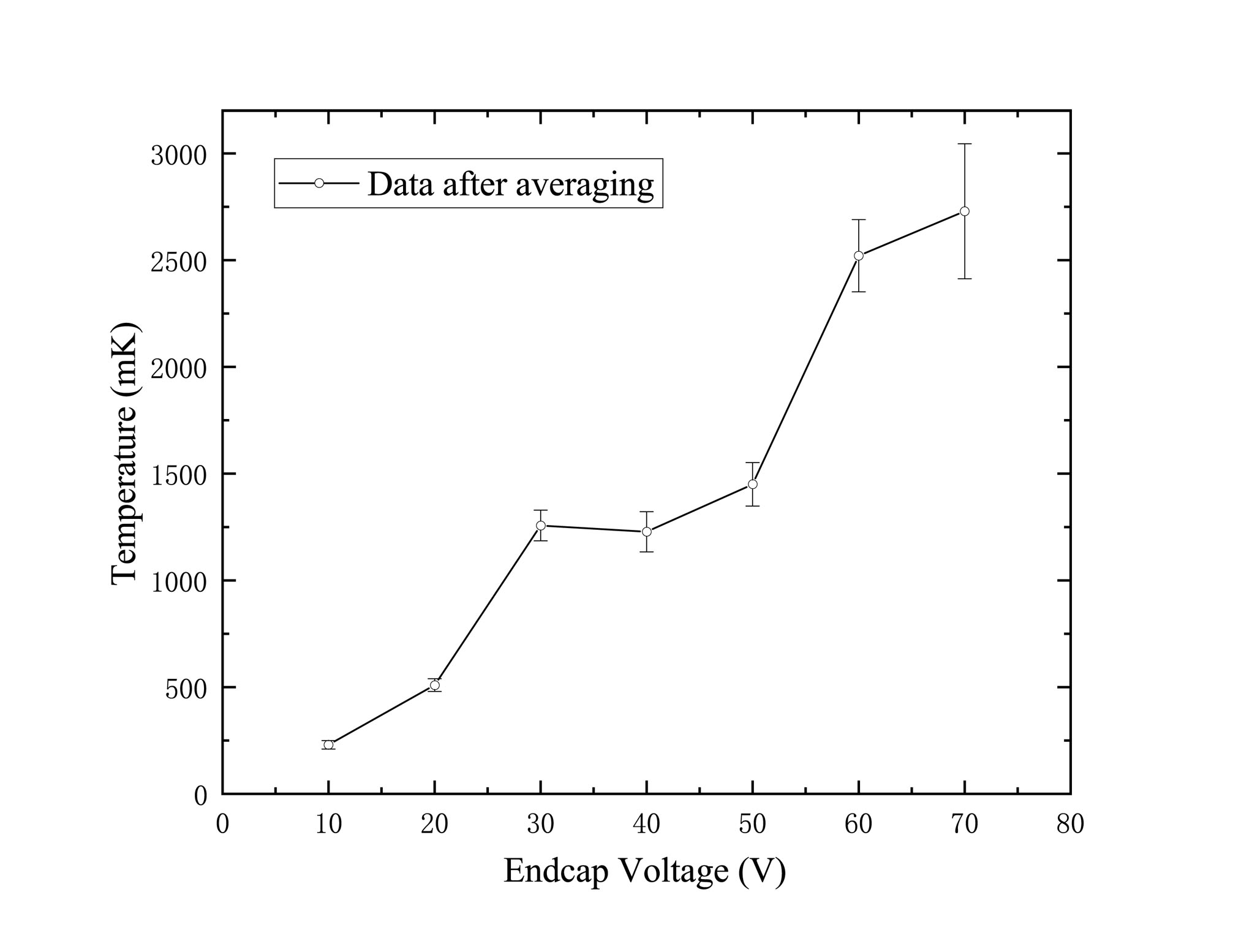}
}
\caption{EC voltage affects the sympathetic cooling efficiency; the RF voltage is fixed at 275 V while changing the EC voltage. Each experimental data point is the average of three measurements.}
\label{fig:EC} 
\end{figure}

Another experimental parameter is the endcap voltage but has no direct influence on the temperature of ions. Nevertheless, as the endcap voltage increases, the ions are squeezed towards the trap axis, and thus more ions will deviate from the trap center, causing an increase in RF heating effects. We measure the temperature of the sympathetic cooled $^{113}$Cd$^+$ ions while changing the endcap voltage (Fig.~\ref{fig:EC}). We find three distinct stages between endcap voltages 30 V--50 V. This includes a sharp rise in $^{113}$Cd$^+$ temperature, which may indicate a phase transition of $^{113}$Cd$^+$ ions into a cloud state. When the endcap voltage is above 60 V, the $^{40}$Ca$^+$ ions cannot be maintained in the liquid phase, and they also cause a sharp increase in temperature of the $^{113}$Cd$^+$ ions.

According to Ref.~\cite{YNZuo-APL-2019}, there is a complicated relationship between the ion temperature and the RF or EC voltage because a decrease in ion number can strongly influence the ion temperature. In this experiment, ions are more stably confined in the ion trap, and hence the number of ions barely decreases. Without the influence of ion number, the relationship between ion temperature and the RF or EC voltage is more intuitive. The experimental results are simply understood from RF heating effects.

We were able to obtain a set of optimized experimental parameters for the $^{40}$Ca$^+$ sympathetic cooling $^{113}$Cd$^+$ scheme, the RF and EC voltage being 250 V and 10 V, respectively. At these parameter settings, the temperature of the $^{113}$Cd$^+$ ions was measured to be 90(10) mK.

\begin{figure}
\centering
\resizebox{0.4\textwidth}{!}{
\includegraphics{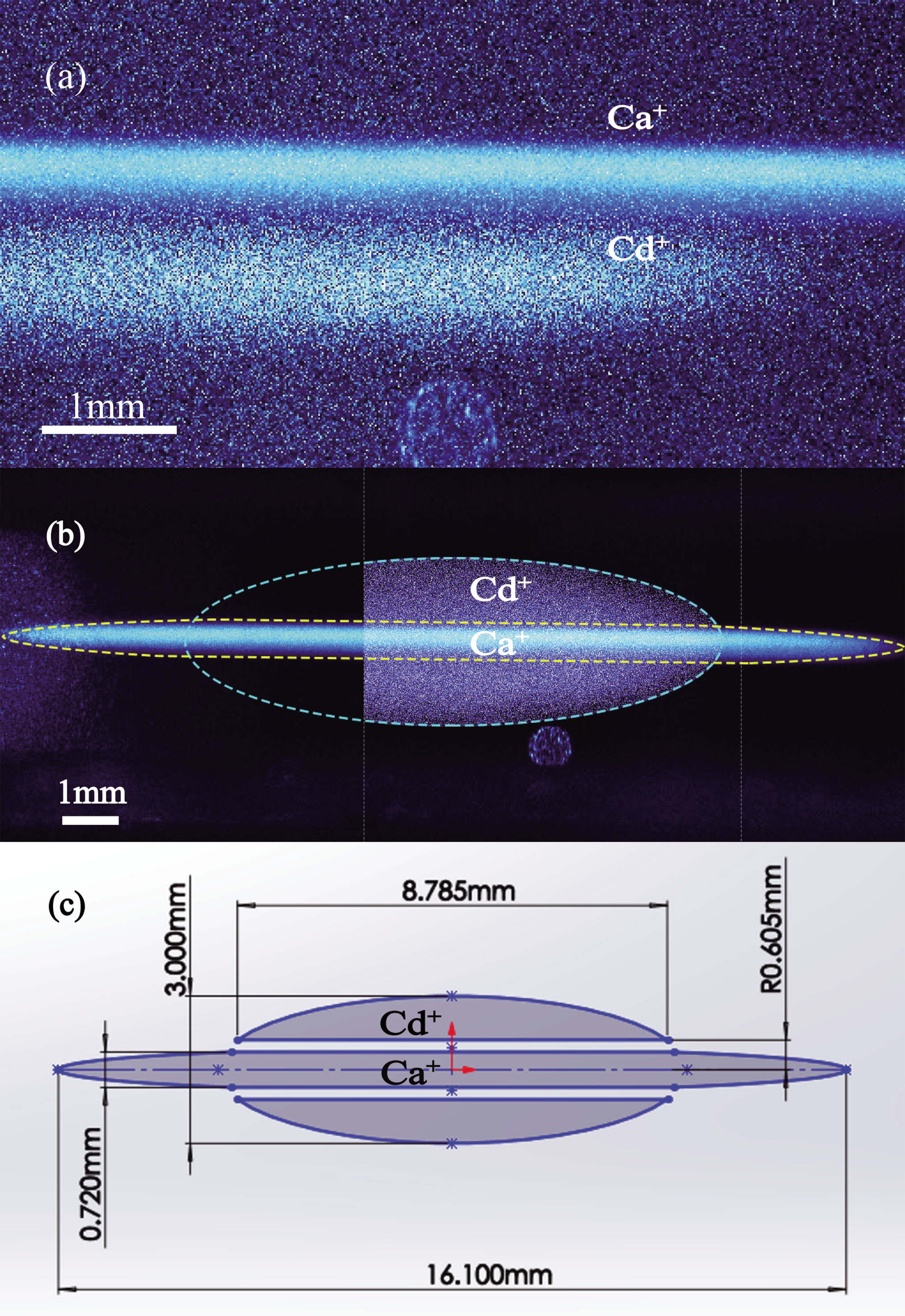}
}
\caption{(a) EMCCD images of the dual-ion system for a 0.3 s exposure time; (b) combined images for Cd$^+$ and Ca$^+$, the Cd$^+$ image being the combination of the seven images from top and bottom, and the Ca$^+$ image being a combination of the three images from left to right (dashed lines represent fitted ion boundaries); (c) fitted dimensions of the Ca$^+$ and Cd$^+$ ion crystals.}
\label{fig:cdca} 
\end{figure}

To find the total number of both ion species in the ion trap, we need to know the volume and density of each species. According to Ref.~\cite{Hornekr-PRL-2001,Wineland-PCCSHCIW-1987}, the configurations of the inner and outer layer of ions are cylindrical and ellipsoidal, and the gap between the inner surface of the ellipsoid shell and the outer surface of the inner cylinder is proportional to the square root of the mass ratio. We use these two configurations to fit the dual-ion images taken with the EMCCD [Fig.~\ref{fig:cdca}(b)]. From the fitted dimension of the Ca$^+$ and Cd$^+$ ion clouds [Fig.~\ref{fig:cdca}(c)], the total ion volume was calculated to be 5.3 mm$^3$ and 34.7 mm$^3$ for $^{40}$Ca$^+$ and $^{113}$Cd$^+$, respectively. In contrast to results in Ref.~\cite{YNZuo-APL-2019}, we see from Fig.~\ref{fig:cdca}(a) and (b) that, in this experimental scheme, there are nearly no dark ions and the ion signals are much more stable, indicating that using $^{40}$Ca$^+$ as coolant to sympathetically cool $^{113}$Cd$^+$ is more efficient and stable.

For ions trapped at low temperature, their density is usually estimated using the zero-temperature approximation \cite{Wineland-PCCSHCIW-1987,Hornekr-PRL-2001},
\begin{equation}\label{eq:n}
n=\frac{\varepsilon_0 V_{RF}^2}{M\Omega^2 r_0^4} ,
\end{equation}
where $\varepsilon_0$ denotes the permittivity of vacuum, $V_{RF}$ the amplitude of RF voltage, $M$ the ion mass, $\Omega$ the trap driving frequency, and $r_0$ the minimum distance from the trap center axis to the surfaces of the electrodes. In this experiment, $V_{RF}=250$V, $\Omega=2\pi\times2.076$ MHz. The number density for $^{113}$Cd$^+$ and for $^{40}$Ca$^+$ were calculated as $n_{Cd}=1.2\times 10^{13}$ m$^{-3}$ and $n_{Ca}=3.3\times10^{13}$ m$^{-3}$, respectively. Therefore, the total number of $^{40}$Ca$^+$ and $^{113}$Cd$^+$ ions in the ion trap are $N_{Cd}=4.2\times10^5$ and $N_{Ca}=1.7\times10^5$, respectively.

\begin{figure}
\centering
\resizebox{0.5\textwidth}{!}{
\includegraphics{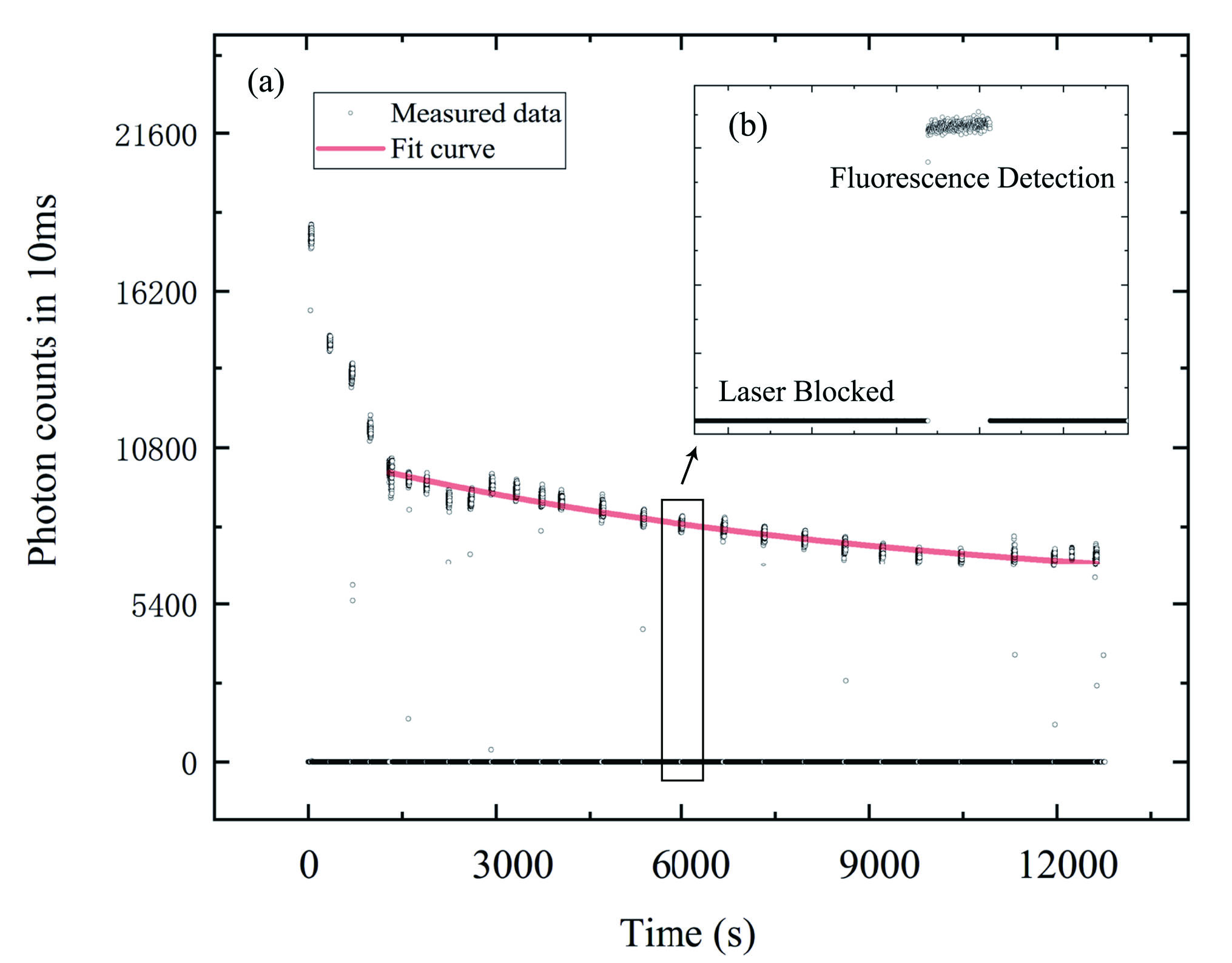}
}
\caption{(a) Ion-loss measurement results fitted with an exponential decay curve. The fitting was started after 25 min of data taking. (b) Ion-loss measurement methods.}
\label{fig:shouming} 
\end{figure}

To analysis the stability of trapping this dual-ion system qualitatively, we measured the loss of $^{113}$Cd$^+$ ions from the ion trap. With the RF voltage and EC voltage fixed, and the power and frequency of the lasers stabilized, we assume the PMT counts to be proportional to the ion number. To avoid laser cooling effects during detection, the detection laser is blocked for 10 min after each 20 s measurement [Fig.~\ref{fig:shouming}(b)]. The ion-loss measurement results are shown in Fig.~\ref{fig:shouming}(a); the counts from background fluorescence have been subtracted. From Fig.~\ref{fig:shouming}(a), we see that the PMT counts decrease rapidly at the beginning. This is because when the number of ions is large, more ions are located at positions far from the center of the trap well and cause confinement of these ions at the edge of the potential well that are not very stable. After about 25 min, the trapping of $^{113}$Cd$^+$ becomes stable, and the rate of ion loss becomes very slow. The fitted curve is an exponential decay curve, and the time constant is about 84 h. After 3.3 h, the PMT count decreased to 40\% (from 18306$\rightarrow$7021), indicating that there are still $1.7\times10^5 $ $^{113}$Cd$^+$ ions in the ion trap. This suffices to provide a good SNR for the microwave ion clock. The measurement of the ion number loss shows that under sympathetic cooling this dual-ion system is very stably within the ion trap. Low-temperature stable trapping of $^{113}$Cd$^+$ is paramount for further developments to microwave ion clocks.

\section{Systematic Shifts and Stability Estimation}
\subsection{Second-order Doppler Shift}
Compared with traditional laser cooling, the sympathetic cooling scheme has many advantages in the application to microwave ion clocks. During pumping and detecting, the large ion clouds are strongly heated by RF heating effects. While the sympathetic cooling scheme is applied, the ions are kept at a low temperature and therefore the uncertainty incurred from SODSs decreases significantly. With the influence of the external field diminished by improving shielding techniques, the SODS has become the fundamental limitation to better accuracy of the microwave ion frequency standard. This highlights the huge potential of applying sympathetic cooling technology in establishing a microwave ion frequency standard.

The estimation of the SODSs for an ion cloud trapped in a Paul trap usually begins with the adoption of a simple model based on the Boltzmann distribution. This simple model is effective at estimating the SODS of ion clouds at high temperature of order of several hundred Kelvin \cite{Cutler-APB-1986,Prestage-IFCS-1999}. For trapped ions at low temperature, the correlated strength is estimated using coupling parameter $\Gamma=q^2/(a k_B T)$, where $a$ denotes the Wigner--Seitz radius $4\pi n a^3/3=1$ \cite{Dubin-RMP-1999}. In our experiment, the temperature of the Cd$^+$ ions is around 90 mK, the density is about $1.2\times10^{13}$m$^{-3}$, and the corresponding coupling parameter is $33$. Under these conditions, the system begins to exhibit local-order characteristics of a fluid \cite{Dubin-RMP-1999}, and there are strong correlations between trapped ions. That is, the SODS of trapped ions at low temperature are obviously different from those of ions at high temperatures. The SODSs of this strongly correlated system need further investigation.

Fortunately, we can roughly estimate SODSs using the equipartition theorem \cite{Berkeland-JAP-1998}, assuming no micromotion along the axial direction of the Paul trap. The SODSs for the trapped ions are roughly estimated to be
\begin{eqnarray}
\frac{\Delta \nu_{\rm{SODS}}}{\nu_{\rm{clock}}}
&=&-\frac{1}{2}\frac{<v^2_{s}>+<v^2_{m}>}{c^2}\nonumber\\
&=&-\frac{1}{2 c^2}(\frac{3k_B T}{M}+\frac{2k_B T}{M})\nonumber\\
&=&-\frac{5 k_B T}{2 M c^2},
\end{eqnarray}
where $\nu_{\rm{clock}}$ denotes the ground state hyperfine splitting of $^{113}$Cd$^+$; $<v^2_{s}>$ and $<v^2_{m}>$ denote the root mean square of the secular motion and the micromotion velocity, respectively. For ions at temperature $90(10)$ mK, the SODS is estimated to be $-1.84(0.2)\times10^{-16}$.

\subsection{AC Stark Shift}
In the sympathetic cooling scheme, the target ions must always be kept in a low-temperature state through the Coulomb interaction, and hence the lasers of the coolant ion must always be kept on. In the experimental scheme of Ca$^+$ sympathetic cooling of Cd$^+$, the Ca$^+$ cooling laser at wavelength 397 nm and the repumping laser at wavelength 866 nm remain on throughout the experiment. Therefore, during Cd$^+$ hyperfine splitting microwave interrogation, these two wavelengths of the lasers and ion fluorescence inevitably shift the clock transition. For a high-accuracy microwave frequency standard, the ac Stark frequency shift induced by the coolant ions needs to be carefully estimated.

Under the influence of light, each clock level is perturbed. The clock transition frequency is modified by the difference in the perturbed energies,
\begin{eqnarray}
\delta\nu^{\rm{Stark}}(\lambda_L)=\delta\alpha(\lambda_L)\cdot I,
\end{eqnarray}
where $\lambda_L$ denotes the perturbed light wavelength, $\delta\nu^{\rm{Stark}}(\lambda_L)$ the ac Stark shift at wavelength $\lambda_L$, $\delta\alpha(\lambda_L)$ the difference in ac Stark polarizabilities of the hyperfine states with F=1 and F=0, and $I$ the perturbed light intensity.

The light that shifts the ion energy level consists of two contributions, one from the cooling and repumping lasers of the coolant ions, and the other from the fluorescence of coolant ions. In Doppler cooling, the light intensity of the lasers is typically set to twice the saturation light intensity of the corresponding transition. For the Ca$^+$ 397 nm and 866 nm transitions, the saturation light intensities are $4.663\times10^{-2}$ W/cm$^2$ and $3.4\times10^{-4}$ W/cm$^2$ \cite{Splatt-phd-2009}, and the corresponding laser intensity $I_{\rm{laser}}$ are $9.3(0.9)\times10^{-2}$ W/cm$^2$ and $6.8(0.7)\times10^{-4}$ W/cm$^2$ with $10\%$ power fluctuations.

The fluorescence of coolant ions is calculated from the spontaneous emission (SE). The SE scattering rate $R$ in a laser-cooling process is estimated from \cite{Foot-book-2005}
\begin{equation}\label{eq:scatt}
R=\frac{\Gamma}{2}\frac{\Omega^2/2}{\delta^2+\Omega^2/2+\Gamma^2/4},
\end{equation}
where $\Gamma$ denotes the transition natural linewidth, $\delta$ the frequency detuning from resonance, and $\Omega$ the Rabi frequency. The Rabi frequency and saturation intensity are related by $I_{\rm{laser}}/I_{sat}=2\Omega^2/\Gamma^2$. For the Ca$^+$ cooling (397 nm, $\Gamma=2\pi\times20.67$ MHz, $\delta=-\Gamma/2$) and repumping (866 nm, $\Gamma=2\pi\times1.69$ MHz, $\delta=0$) transitions, the scattering rates are $3.2\times10^7$s$^{-1}$ and $3.5\times10^6 $ s$^{-1}$, respectively.

The intensity of SE is calculated using $I=NRh\nu/A$, where $N=1.7\times10^5$ is the number of Ca$^+$ ions, $h$ the Planck constant, $\nu$ the frequency of fluorescence, and $A$ the luminous area. We assume that the luminous area is equal to the external surface area of Ca$^+$ ion crystal and all the fluorescence influences the clock transition (as an upper limit). We find $A$ is 31.7 mm$^2$ (see Fig.~\ref{fig:cdca}). The SE intensity $I_{\rm{\rm{fluo.}}}$ of the Ca$^+$ cooling and repumping transitions are estimated to be less than $8.6\times10^{-6}$ W/cm$^2$ and $4.3\times10^{-7}$ W/cm$^2$, respectively.

The ac Stark polarizabilities $\delta\alpha$ at wavelengths 397 nm and 866 nm were analyzed from theory based on second- and third-order perturbation techniques, taking into account the hyperfine interaction, which has been described in detail in Ref.~\cite{Rosenbusch-PRA-2009}. The calculations were performed using a high-accuracy \textit{ab initio} approach that included the important many-body and relativistic effects \cite{Angstmann-PRA-2006}. The calculation results of the ac Stark shifts are listed in Table \ref{Tab:acstark}, along with the relative ac Stark shift $\delta\nu^{\rm{Stark}}/\nu_{\rm{clock}}$.

\begin{table}[t]
\caption{Calculation results of the ac Stark shifts for different wavelengths.} \label{Tab:acstark}
{\setlength{\tabcolsep}{4pt}
\begin{tabular}{l|rr}\hline\hline
$\lambda_L$[nm] & 397 & 866 \\
$\delta\alpha$ [Hz/(W/cm$^2$)] & $8.75\times10^{-6}$ & $4.54\times10^{-6}$\\
$I_{\rm{laser}}$ [W/cm$^2$] &$9.3(0.9)\times10^{-2}$&$6.8(0.7)\times10^{-4}$\\
$\delta\nu^{\rm{Stark}}_{\rm{laser}}$[Hz] &$8.1(0.8)\times10^{-7}$
&$3.1(0.3)\times10^{-9}$\\
$\delta\nu^{\rm{Stark}}_{\rm{laser}}/\nu_{\rm{clock}}$ &$5.4(0.5)\times10^{-17}$
&$2.0(0.2)\times10^{-19}$\\
$I_{\rm{fluo.}}$ [W/cm$^2$] &$<8.6\times10^{-6}$&$<4.3\times10^{-7}$\\
$\delta\nu^{\rm{Stark}}_{\rm{fluo.}}$[Hz] &$<8\times10^{-11}$&$<2\times10^{-12}$\\
$\delta\nu^{\rm{Stark}}_{\rm{fluo.}}/\nu_{\rm{clock}}$ & $<6(3)\times10^{-21}$ & $<2(1)\times10^{-22}$\\
\hline\hline
\end{tabular}}
\end{table} 

The results show that the magnitude and uncertainty of the ac Stark shifts were down to $10^{-17}$ and $10^{-18}$ levels. This is because the two laser frequencies of Ca$^+$ are detuned far from all possible transition frequencies of the Cd$^+$ energy levels. The small shifts indicate that, with the Ca$^+$ sympathetic cooling Cd$^+$ scheme, we no longer need to worry about the influence of the ac Stark shift.

\subsection{Dick Effect}
The Dick effect arising from phase noise properties of the local oscillator and the unavoidable dead time, is one of the main limiting factors to short-term frequency stability. The sympathetic cooling scheme offers an effective way to reduce the influence of the Dick effect on stability. The Dick-effect-limited Allan deviation may be expressed as \cite{JWZhang-CPL-2015,Santarelli-ITUFFC-1998}
\begin{equation}\label{eq:dick}
\sigma^{Dick}_y(\tau)=[\frac{1}{\tau}\sum\limits_{m=1}^{\infty}
(\frac{g_{ms}^2+g_{mc}^2}{g_{0}^2})S^f_y(\frac{m}{2T_c})]^{-1/2},
\end{equation}
where $\tau$ denotes the sampling time, $m$ the positive integer, $T_c$ the total time of one measurement cycle, and $S^f_y(m/2T_c)$ the one-sided power spectral density of the relative frequency fluctuations of a free running interrogation oscillator at Fourier frequencies $m/2T_c$. The parameters $g_{ms}$, $g_{mc}$, and $g_{0}$ are defined as
\begin{eqnarray}
g_{ms}&=&\frac{1}{2T_c}\int_{0}^{2T_c} \ g(\theta)sin(\frac{\pi m \theta}{2T_c})\, d\theta,\nonumber\\
g_{mc}&=&\frac{1}{2T_c}\int_{0}^{2T_c} \ g(\theta)cos(\frac{\pi m \theta}{2T_c})\, d\theta,\nonumber\\\
g_{0}&=&\frac{1}{2T_c}\int_{0}^{2T_c} \ g(\theta)\, d\theta,
\end{eqnarray}
where $g(\theta)$ denotes the Ramsey interrogation sensitivity function \cite{Santarelli-ITUFFC-1998}.

The time sequences of the laser-cooled and sympathetic-cooled $^{113}$Cd$^+$ microwave atomic clocks are presented in Table \ref{Tab:time}. The cycling transition $5s~^2S_{1/2}\ (F=1\ m_F=1)\rightarrow 5p~^2P_{3/2}\ (F=2\ m_F=2)$ is used for laser cooling, the transition $5s~^2S_{1/2}\ (F=1)\rightarrow5p~^2P_{3/2}\ (F=1)$ is used for pump ions into state $5s~^2S_{1/2}\ (F=0\ m_F=0)$, and the hyperfine transition $5s~^2S_{1/2}\ (F=0\ m_F=0)\leftrightarrow5s~^2S_{1/2}\ (F=1\ m_F=0)$ is the clock transition (see Fig.~\ref{fig:level}). By using the sympathetic cooling scheme, there is no need for an extra cooling procedure. The cycle time is then reduced from 1830 ms to 830 ms. Using the same local oscillator as mention in \cite{JWZhang-CPL-2015}, the Dick-effect-limited Allan deviation was reduced from $3.8\times10^{-14}/\sqrt{\tau}$ to $1.8\times10^{-14}/\sqrt{\tau}$, with the frequency stability limit being reduced by half.

\begin{table}[t]
\caption{Time sequences of the laser-cooled and sympathetic-cooled $^{113}$Cd$^+$ microwave atomic clocks.} \label{Tab:time}
{\setlength{\tabcolsep}{12pt}
\begin{tabular}{l|ll}\hline\hline
Procedure & $T_{\rm{Laser}}$ (ms) & $T_{\rm{Symp.}}$ (ms)\\\hline 
Cooling &1000 & 0\\
Pumping &50 & 50\\
$\pi/2$ Pulse &15 & 15 \\
Free evolution & 500 & 500\\
$\pi/2$ Pulse &15 & 15\\
Detecting &250 & 250\\
\\
Dead time &1300 &300\\
Cycle time &1830 &830\\
\hline\hline
\end{tabular}}
\end{table}

The fundamental fractional systematic frequency shifts and stability of the laser-cooled \cite{KMiao-OL-2015} and sympathetic-cooled Cadmium ion microwave clock systems are compared in Table \ref{Tab:sum}. The total uncertainty associated with these systematic shifts is reduced by three orders of magnitude to the $10^{-17}$ level, indicating great potential for developing ultra-high accuracy microwave ion clocks.

\begin{table}[t]
\caption{Fractional systematic shifts and stability of laser-cooled and sympathetic-cooled $^{113}$Cd$^+$ microwave atomic clock.} \label{Tab:sum}
{\setlength{\tabcolsep}{4pt}
\begin{tabular}{l|rr}\hline\hline
Item & Laser-cooled & Symp.-cooled \\\hline 
Second-order &$-1.8(0.3)\times10^{-14}$ & $-1.84(0.2)\times10^{-16}$\\
Doppler&&\\
ac Stark shift &0 & $5.4(0.5)\times10^{-17}$\\
(Laser)&&\\
ac Stark shift & 0 & $<6(1)\times10^{-21}$\\
(Fluorescence)&&\\
Dick-effect-limited & $3.8\times10^{-14}/\sqrt{\tau}$ &$1.8\times10^{-14}/\sqrt{\tau}$\\
Allan Deviation&&\\
\hline\hline
\end{tabular}}
\end{table}

\section{Conclusions}
We achieved long-term low-temperature trapping of a large number of sympathetically cooled $^{113}$Cd$^+$ ions using laser-cooled $^{40}$Ca$^+$ as coolant. RF trapping was very stable for this Ca$^+$-Cd$^+$ system. After three hours in the trap, up to $10^5$ Cd$^+$ ions were still confined. The temperature of the $^{113}$Cd$^+$ ions was as low as 90 mK, which aided the suppression of the uncertainty associated with the SODS. Using the equipartition theorem, this uncertainty was roughly estimated to be at the $10^{-17}$ level. The ac Stark shifts incurred by the Ca$^+$ cooling lasers and fluorescence were carefully estimated. The ac Stark polarizabilities were calculated by high-accuracy \textit{ab initio} calculations. Because the frequency differences between the two laser frequencies of Ca$^+$ and all possible transitions of Cd$^+$ are large, the magnitude and uncertainty were down to $10^{-17}$ and $10^{-18}$ levels, and hence can be ignored in these circumstances. The results indicate that using Ca$^+$ sympathetic cooling of Cd$^+$ is expected to be an effective experimental method to improve the accuracy of the Cadmium-ion microwave frequency standard to order $10^{-16}$. In this sympathetic cooling scheme, one cooling procedure is no longer needed. By decreasing dead time, the Dick-effect-limited Allan deviation is also reduced. All of the experimental results and theoretical analysis show that a microwave clock based on sympathetically cooled Cd$^+$ ions holds promise to attain ultrahigh frequency accuracy and stability.

\section*{Acknowledgements}
We would like to thank Z. Y. Peng and K. Beloy for their helpful discussions and suggestions. This work is supported by the National Key Research and Development Program of China (No. 2016YFA0302100), and the Beijing Natural Science Foundation (1202011). Y. M. Yu acknowledges the National Natural Science Foundation of China (No. 11874064), and the Strategic Priority and Research Program of the Chinese Academy of Sciences (No. XDB21030300). J. Z. Han thanks S. C. Wang for image processing.

\end{document}